\documentclass{sig-alternate}

\usepackage{enumitem}
\usepackage[hidelinks]{hyperref}
\usepackage{color}
\usepackage{xcolor,colortbl}
\usepackage{multirow}
\usepackage{hhline}
\usepackage{etoolbox}

\hypersetup{colorlinks=false}
 
\definecolor{Gray}{gray}{0.90}
\newcolumntype{L}{>{\centering\arraybackslash}m{15mm}}

\makeatletter
\def\@copyrightspace{\relax}
\makeatother

\begin{document}


\title{An initial performance review of software components for a heterogeneous computing platform}
%
%
%
%

\numberofauthors{2} 
%
\author{
%
%
\alignauthor
Ivan \v{S}vogor\\
       \affaddr{University of Zagreb}\\
       \affaddr{Faculty of Organization and Informatics}\\
       \affaddr{Pavlinska 2, 42000 Vara\v{z}din, Croatia}\\
       \email{isvogor@foi.hr}
}



\maketitle

\begin{abstract}

The design of embedded systems is a complex activity that involves a lot of decisions. With high performance demands of present day usage scenarios and software, they often involve energy hungry state-of-the-art computing units. While focusing on power consumption of computing units, the physical properties of software are often ignored. Recently, there has been a growing interest to quantify and model the physical footprint of software (e.g. consumed power, generated heat, execution time, etc.), and a component based approach facilitates methods for describing such properties. Based on these, software architects can make energy-efficient software design solutions. This paper presents power consumption and execution time profiling of a component software that can be allocated on heterogenoeus computing units (CPU, GPU, FPGA) of a tracked robot.

\end{abstract}




\section{Introduction}

The design of present day embedded systems and the accompanying software is becoming ever more complex, and with an emergence of specialized computing units for accelerating certain operations, this trend is growing. Along with benefits, this also carries some drawbacks, i.e. additional complexity which generates a variety of (side--)effects that are hard to ignore while designing software. From the cyber-physical systems theory perspective~\cite{Wolf2009, Poovendran2010}, software is characterized by physical properties, which are apparent in its requirement and consumption of resources, e.g. time to execute, the power consumed, the generated heat, etc.~\cite{Wolf2012}. For software architects this presents additional considerations in the software design process. Component based software engineering facilitates techniques for expressing these system characteristics in the form of extra--functional properties~\cite{Member2011}, based on which software architects perform software design decisions. 

This paper presents the collection process and analysis of extra--functional properties for a component based software architecture deployable on multiple computing units within a heterogeneous computing environment. The focus is on execution time which reflects the system performance and power consumption which is increasingly important in both everyday computing and remote--hazardous environments.  

\subsection{Background}

This research is the continuation of an effort to create a decision making framework for allocation of component based software on a heterogeneous computing platform~\cite{Svogor2013a}. The allocation decision is based on the specified extra--functional properties by the software architect. Assuming that one software component can be allocated on different computing units within a computing platform, software architect is faced with a large decision space $m^n$ (where $m$ is the number of computing units and $n$ is the number of software components). Therefore, we have created a multi--criterion framework that presents a software architect with an optimal allocation. However, it was only tested on computer generated data, so in order to improve the reliability of the framework, this paper provides information about collecting real--world data that will be used to fine tune it. 

\subsection{System description}

The software components profiled in this paper are the part of the software architecture, which is designed for a tracked robot, carrying a heterogeneous computing platform. The platform is composed of commercial off--the--shelf components (COTS), selected in a way to minimize power consumption. It consists of a multicore CPU (Intel i3-3240), a GPU (Radeon HD7750 1GB low profile) and an FPGA (Xilinx Spartan 6)\footnote{LogiPi board, a Kickstarter FPGA project for RaspberryPi -- \url{http://valentfx.com/logi-pi/}}. These components are hosted by a mini-ITX Asus P8H61-I motherboard with Kingston HyperX SSD ($256GB$) and RAM ($8GB$). For computer vision it currently uses Microsoft Kinect, and the operating system is Linux Ubuntu 14.10 LTS. 

\begin{figure}[h]
    \centering
		\includegraphics[trim = 5mm 5mm 5mm 5mm, clip, width=0.90\columnwidth]{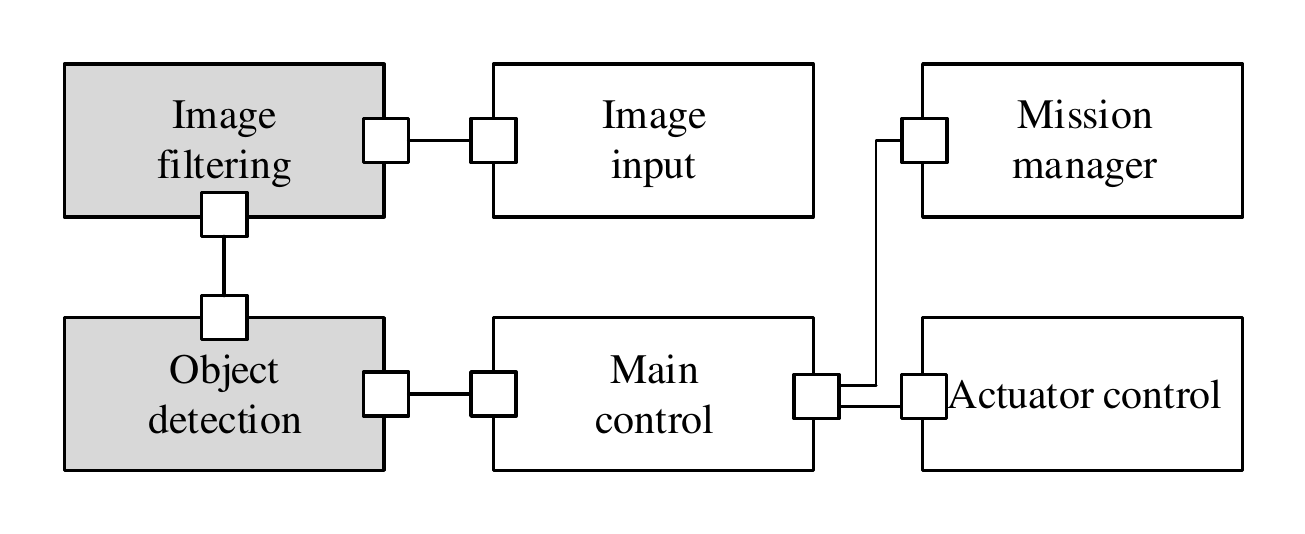}
    \caption{Simplified software architecture layout}
    \label{fig:swArch}
\end{figure}

\noindent \autoref{fig:swArch} presents an overview of the software components. In this paper, only two components are analyzed; \textsl{Image filtering} and \textsl{Object detection}. The Image filtering component can be allocated on  (i.e. is implemented for) a CPU, a GPU and an FPGA, while the Object detection component can be allocated on a CPU and a GPU. 
The following section presents the collection of data for an average execution time and an average power consumption of these components. The collected data is intended to be used by software architects to make decisions about software component allocation.

\section{Measurement and results}

The Image filtering and the Object detection components are implemented in Java, however both use Java native interface (JNI) to access OpenCL (AMD's C++ implementation) and OpenCV (C++ implementation). Image filtering component provides OpenCL kernels for Gaussian blur, Sobel, Erode, Dilate and Hysteresis filters, while Object detection component uses OpenCL extensions for OpenCV to detect objects (using a custom Haar's classifier for arrow detection). For the FPGA, IP cores for Gaussian blur, Sobel, Erode, Dilate and Hysteresis filters are provided by LogiPi.

\subsection{Power consumption}

The method used to profile the power consumption of the software components is similar to the one presented by Collange et al.~\cite{Collange2009}. While measuring the voltage and current of the system\footnote{With UNI-T UT151C multimeter}, power consumption of the CPU was monitored by Intel's Power Gadget tool. Repeated measurements have shown that while idling the average power consumption of the entire system was $32.38W$, however with the GPU removed it was only $13.59W$. Hence, the average power consumption of the idling GPU was $18.78W$.

\subsubsection*{Image filtering component}

To acquire the average power consumption of a component, the functions from the components were called 25 times, while  measuring voltage and current. 25 repetitions are rounded up from 21, necessary to statistically achieve confidence level of 95\% with 10\% error margin.
 \autoref{tab:filteringPower} shows the results; average power consumption of the system, SD, variance and the average power consumption of the active software component, i.e. a raise of the average power consumption while the component was active. 

\begin{table}[t!]
	\centering
		\caption{Filtering component power consumption (per call)}
		\resizebox{\columnwidth}{!}{%
		\begin{tabular}{ |l|l|l|L|l|l|L| }
		\hhline{*{2}{~}*{5}{|-}}
			\multicolumn{2}{c|}{} & Filters & Avg.pwr. system [W] & SD ($\sigma$) & Var ($\sigma^{2}$) & Avg.pwr. sw.comp. [W] \cellcolor{Gray} \\ \hline\hline
			\multirow{7}{*}{\rotatebox[origin=c]{90}{GPU}}	 	
			&	\multirow{3}{*}{\rotatebox[origin=c]{90}{Small}} 	
			& 	Gauss									&	35.8794	& 1.3739	& 1.8877 & 3.5006 \cellcolor{Gray}\\ 
			&	& Gauss, sobel					& 36.7176	& 2.2792	& 5.1950 & 4.3388 \cellcolor{Gray}\\ 
			&	& Gauss, sobel, dilate	&	37.7720	& 3.6222	& 13.1208 & 5.3932\cellcolor{Gray}\\  \hhline{*{1}{|~}*{6}{|-}}
			&	\multirow{4}{*}{\rotatebox[origin=c]{90}{Big}} 	
			& 	Gauss									&	36.5306	& 1.6057	& 2.5784 & 4.1518 \cellcolor{Gray}\\ 
			&	& Gauss, sobel					& 36.7205	& 2.1783	& 4.7451 & 4.3417 \cellcolor{Gray}\\  
			&	& Gauss, sobel, dilate	&	37.9467	& 3.8378	& 14.7292 & 5.5679 \cellcolor{Gray}\\ 
\hhline{*{2}{|~}*{5}{|-}}
			&	&	Sobel (biggest)				&	53.9662	& 7.4789	& 55.9349  & 21.5874 \cellcolor{Gray}\\ \hline \hline 				
			\multirow{7}{*}{\rotatebox[origin=c]{90}{CPU}}	 	
			&	\multirow{3}{*}{\rotatebox[origin=c]{90}{Small}} 	
			& 	Gauss									&	36.5834 &	1.2332 &	1.5209 & 4.2046 \cellcolor{Gray}\\ 
			&	& Gauss, sobel					& 37.1204 &	2.0879 &	4.3596 & 4.7416 \cellcolor{Gray}\\  
			&	& Gauss, sobel, dilate	&	38.0295 &	3.3541 &	11.2501 & 5.6507 \cellcolor{Gray}\\ \hhline{*{1}{|~}*{6}{|-}}
			&	\multirow{4}{*}{\rotatebox[origin=c]{90}{Big}} 	
			& 	Gauss									&	37.0105 &	1.8153 &	3.2956 & 4.6317 \cellcolor{Gray}\\ 
			&	& Gauss, sobel					& 38.1082 &	3.4041 &	11.5883 & 5.7294 \cellcolor{Gray}\\  
			&	& Gauss, sobel, dilate	&	39.3136 &	5.5675 &	30.9974 & 6.9349 \cellcolor{Gray}\\ \hhline{*{2}{|~}*{5}{|-}}		
			&	&	Sobel (biggest)				&	45.1727 &	12.0519 &	145.2498 & 12.7939\cellcolor{Gray}\\ \hline \hline			
			\multirow{3}{*}{\rotatebox[origin=c]{90}{FPGA}}	 	
			&	\multirow{3}{*}{\rotatebox[origin=c]{90}{Small}} 	
			& 	Gauss									&	2.1049 &	0.1003 &	0.0100 & 0.3651 \cellcolor{Gray}\\ 
			&	& Gauss, sobel					& 2.1187 &	0.1134 &	0.0128 & 0.3788 \cellcolor{Gray}\\  
			&	& Gauss, sobel, dilate	&	2.1284 &	0.0986 &	0.0097 & 0.3885 \cellcolor{Gray}\\ \hline			
			\multicolumn{7}{l}{\textsl{* big image: 1280$\times$720px, small image: 320$\times$240 px, biggest: 8192$\times$8192 px }}\\										
		\end{tabular}
		}
	\label{tab:filteringPower}
\end{table}

\begin{figure}[h!]
    \centering
		\includegraphics[trim = 0mm 0mm 0mm 0mm, clip, width=0.95\columnwidth
		]{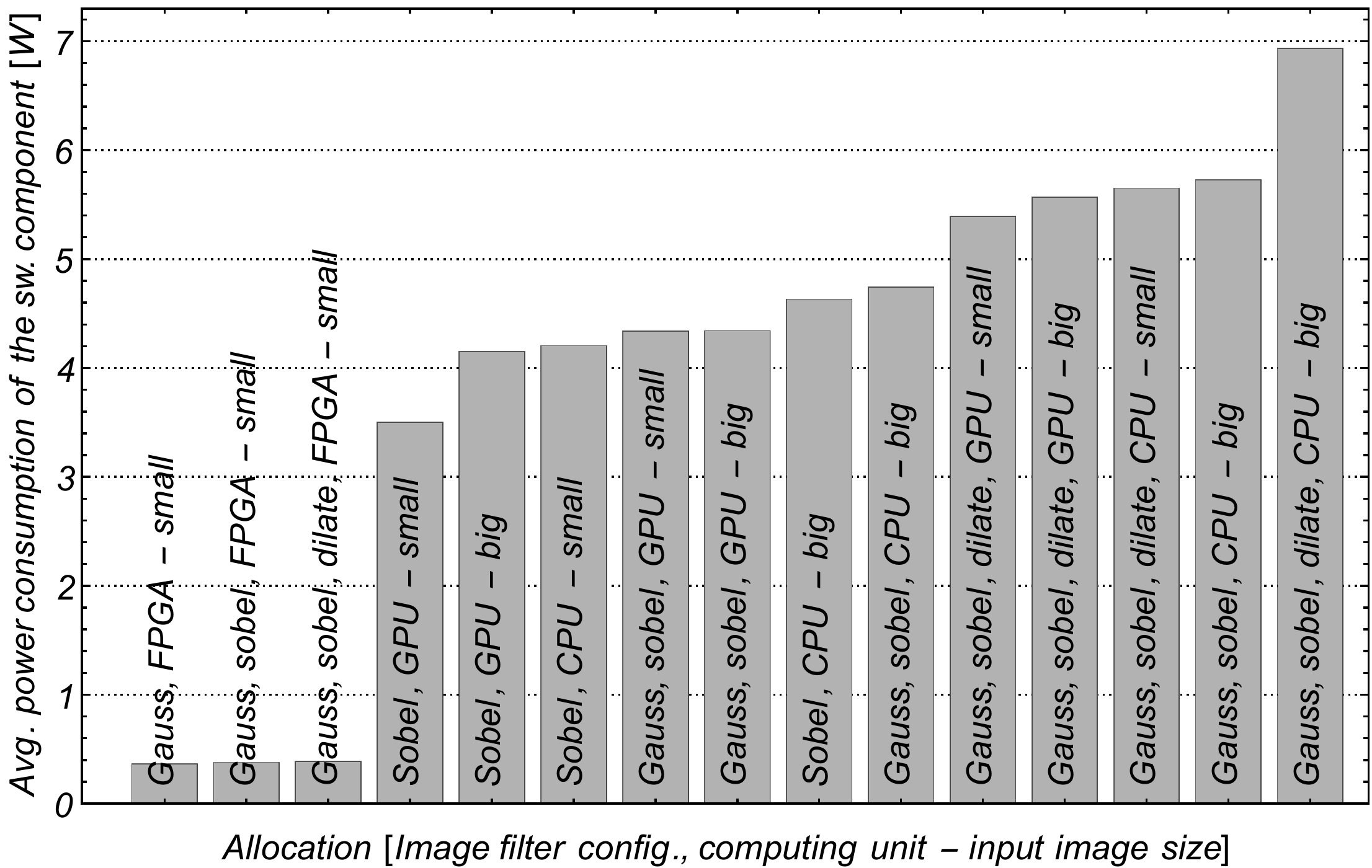}
    \caption{Average power consumption rankings}
    \label{fig:energyFilters}
\end{figure}

\autoref{fig:energyFilters} shows the comparison of an average power consumption for different function calls and different allocations of the Image filtering component. For the FPGA, the IP cores only supported small images, but regardless of this it consumed at least 10 times less power than other options. Surprisingly, while filtering high--resolution images, the CPU consumed the most power (which is related to execution time, addressed in~\autoref{sec:executionTime}). To further test this (not displayed on~\autoref{fig:energyFilters}), both the CPU and the GPU software allocations were presented with a 8192$\times$8192 $px$ (8K) image to perform the Sobel filter. In that case (see ~\autoref{tab:filteringPower}), GPU consumed more power.

\begin{figure}[ht]
    \centering
		\includegraphics[width=\columnwidth]{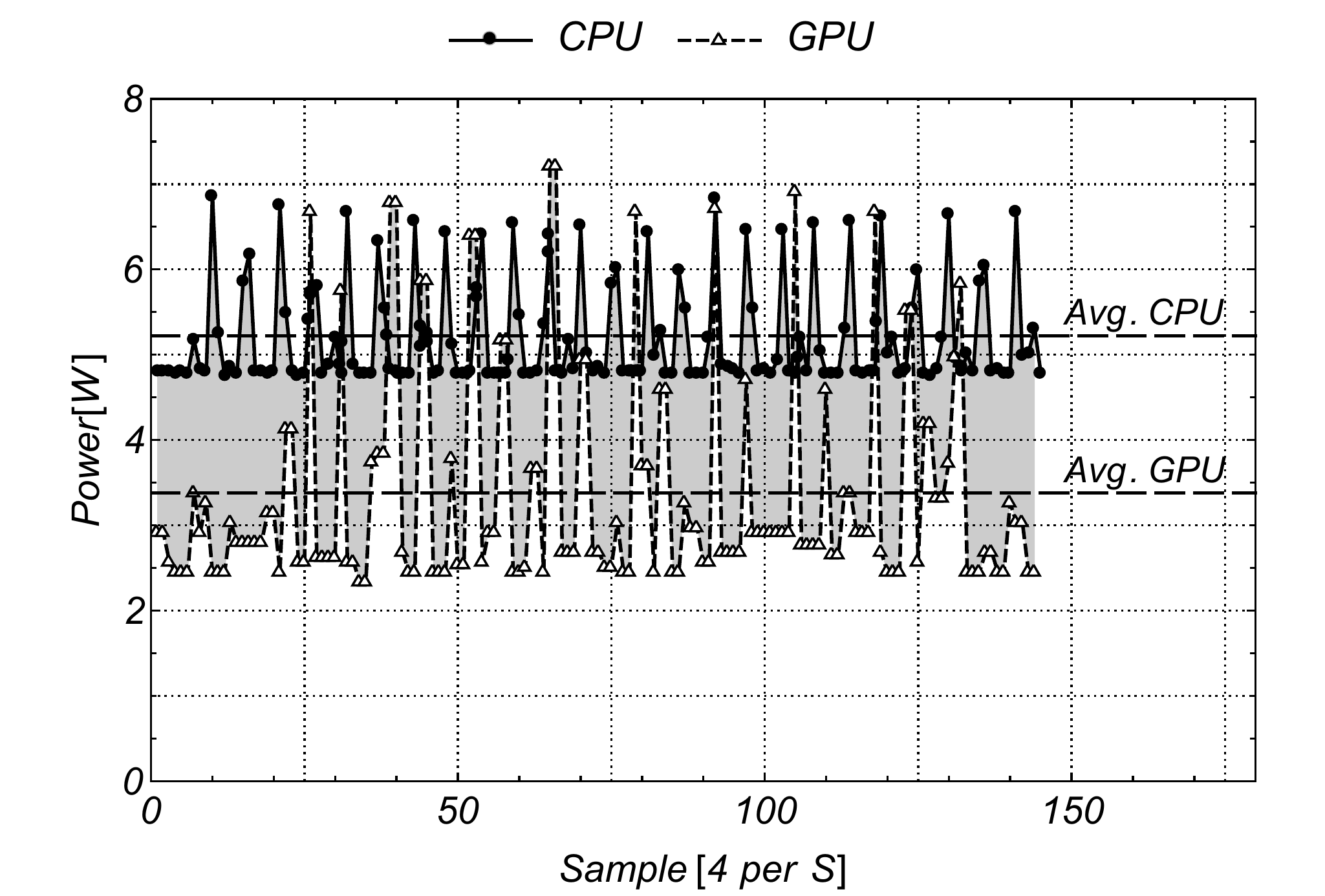}
    \caption{Avg. power consumption of GPU and CPU for Sobel filter at GPU on a small image (25 function calls, reduced by idle power consumption)}
    \label{fig:energyFiltersGraph}
\end{figure}

\noindent An interesting observation is shown in \autoref{fig:energyFiltersGraph}. When allocated on the GPU, the Image filtering component processing small sized images consumes less power than the CPU, which is attributed to data handling operations, as shown by Collange et al.~\cite{Collange2009}. For software architects this means, that while a component is allocated on a GPU, in reality the code cannot be executed without a CPU host. Therefore, components allocated on a GPU will always attribute to power consumption from both a CPU and a GPU. The spikes visible on the graph represent 25 function calls.

\subsubsection*{Object detection component}

\noindent The average power consumption of the Object detection component is shown in \autoref{tab:detectionPower}. Since this measurement was automated, the number of repetitions could be increased, so the measurement consisted of 400 function calls of the Object detection component to achieve statistical confidence level of 99\% with error margin of 2.5\% (rounded up from 315). From the power consumption perspective, even for smaller images, the execution of the component on the CPU consumes less power than the component on the GPU.

\begin{table}[!h]
	\centering
		\caption{Detection component power consumption (per call)}
		\resizebox{0.85\columnwidth}{!}{%
		\begin{tabular}{ |l|l|L|l|l|L| }
		\hhline{*{1}{~}*{5}{|-}}
			\multicolumn{1}{c|}{}& Img. size &  Avg.pwr. system [W] & SD ($\sigma$) & Var ($\sigma^{2}$) & Avg.pwr. sw.comp. [W] \cellcolor{Gray}\\ \hline\hline				
			\multirow{2}{*}{\rotatebox[origin=c]{90}{GPU}} 	
			&	Small					& 58.3868 &	14.7791 &	218.4228 & 26.0080 \cellcolor{Gray}\\ 		
			& Big						&	69.7237 &	16.2128 & 262.8550 & 37.3449 \cellcolor{Gray}\\ \hline \hline
			\multirow{2}{*}{\rotatebox[origin=c]{90}{CPU}} 
			&	Small					& 52.4355 &	3.2249 & 10.4001 & 18.6756 \cellcolor{Gray}\\ 		
			& Big						&	51.0544	& 4.5903 & 21.0713 & 20.0567 \cellcolor{Gray}\\ \hline
			\multicolumn{6}{l}{\textsl{* big image: 1920x1080px, small image: 853x480px}} \\
		\end{tabular}
		}
	\label{tab:detectionPower}
\end{table}

Although the average power consumption by the GPU was $37.32W$ for big images, measurements have shown that for short bursts it peaks at $51.18W$, while for the CPU the peak was at $29.45W$. 

\subsection{Execution time}
\label{sec:executionTime}

\subsubsection*{Image filtering component}

While measuring power consumption, execution time for each function call was recorded by software, so for both components the number of repetitions is the same as for power consumption measurement. 

\begin{table}[t!]
	\centering
		\caption{Filtering component execution times}
		\resizebox{\columnwidth}{!}{%
		\begin{tabular}{ |l|l|l|l|l|l| }
		\hhline{*{2}{~}*{4}{|-}}
			\multicolumn{2}{c|}{} & Filters & Avg.exec.t. [ms] \cellcolor{Gray} & SD ($\sigma$) & Var ($\sigma^{2}$) \\ \hline\hline
			\multirow{7}{*}{\rotatebox[origin=c]{90}{GPU}}	 	
			&	\multirow{3}{*}{\rotatebox[origin=c]{90}{Small}} 	
			& 	Gauss									&	0.4696	\cellcolor{Gray}& 0.0021 	& 0.0000 \\ 
			&	& Gauss, sobel					& 1.1188 	\cellcolor{Gray}& 0.0554	& 0.0030 \\  
			&	& Gauss, sobel, dilate	&	1.7822 	\cellcolor{Gray}&	0.1162 	&	0.0135 \\ \hhline{*{1}{|~}*{5}{|-}}
			&	\multirow{4}{*}{\rotatebox[origin=c]{90}{Big}} 	
			& 	Gauss									&	5.4338 	\cellcolor{Gray}&	0.0035	&	0.0000 \\ 
			&	& Gauss, sobel					& 15.1985	\cellcolor{Gray}& 1.4826	&	2.1981 \\  
			&	& Gauss, sobel, dilate	&	22.7772	\cellcolor{Gray}&	1.5903	&	2.5292 \\ \hhline{*{2}{|~}*{4}{|-}}			
			&	&	Sobel (biggest)				&	115.9208 \cellcolor{Gray}&	0,5215	&	0,2720 \\ \hline \hline 				
			\multirow{7}{*}{\rotatebox[origin=c]{90}{CPU}}	 	
			&	\multirow{3}{*}{\rotatebox[origin=c]{90}{Small}} 	
			& 	Gauss									&	1.1896	\cellcolor{Gray}&	0.2481	&	0.0615 \\ 
			&	& Gauss, sobel					& 3.2682	\cellcolor{Gray}&	0.6001	&	0.3601 \\  
			&	& Gauss, sobel, dilate	&	5.2714	\cellcolor{Gray}&	1.1285	&	1.2736 \\ \hhline{*{1}{|~}*{5}{|-}}
			&	\multirow{4}{*}{\rotatebox[origin=c]{90}{Big}} 	
			& 	Gauss									&	13.8018	\cellcolor{Gray}&	1.2850 &	1.6513 \\ 
			&	& Gauss, sobel					& 32.9411	\cellcolor{Gray}&	2.7509 &	7.5679 \\  
			&	& Gauss, sobel, dilate	&	49.0965	\cellcolor{Gray}&	4.1844 &	17.5100 \\ \hhline{*{2}{|~}*{4}{|-}}				
			&	&	Sobel (biggest)				&	633.4548 \cellcolor{Gray}&	12.8369 &	164.7860 \\ \hline \hline			
			\multirow{3}{*}{\rotatebox[origin=c]{90}{FPGA}}	 	
			&	\multirow{3}{*}{\rotatebox[origin=c]{90}{Small}} 	
			& 	Gauss									&	21.4378 \cellcolor{Gray}&	0.0187 &	0.0003 \\ 
			&	& Gauss, sobel					& 22.2371 \cellcolor{Gray}&	0.0073 &	0.0001 \\  
			&	& Gauss, sobel, dilate	&	21.7476 \cellcolor{Gray}&	0.0233 &	0.0005 \\ \hline			
			\multicolumn{6}{l}{\textsl{* big image: 1280$\times$720px, small image: 320$\times$240px, biggest: 8192$\times$8192 px}} \\										
		\end{tabular}
		}
	\label{tab:filteringExec}
\end{table}

\begin{figure}[h!]
    \centering
		\includegraphics[trim = 0mm 0mm 0mm 0mm, clip, width=0.95\columnwidth]{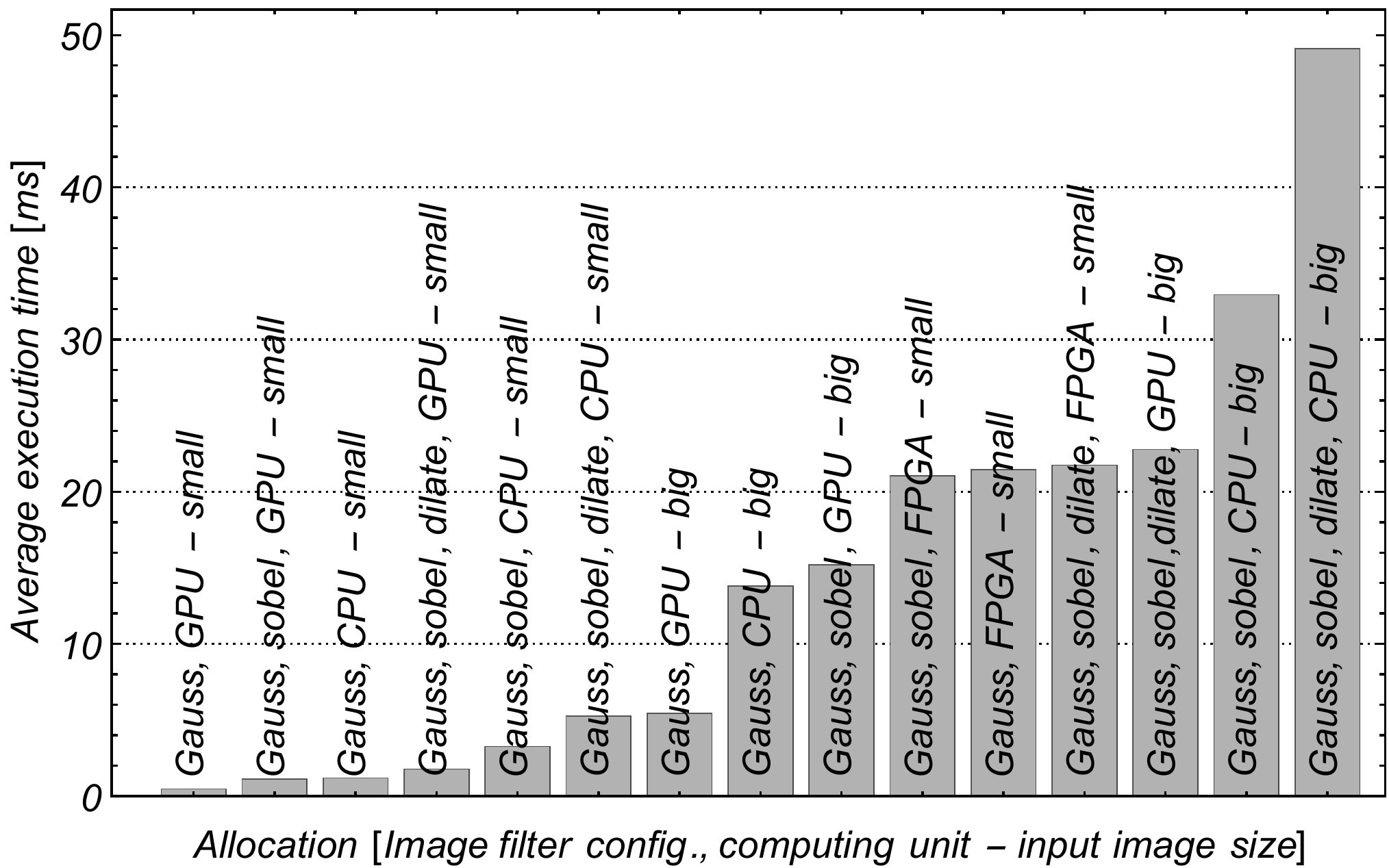}
    \caption{Average execution time ranks}
    \label{fig:execFilters}
\end{figure}

\autoref{tab:filteringExec} shows the results.  For the same filter configuration the average execution time for the component allocated on the GPU is better than the one allocated on the CPU, in all cases, for both small and big images. Nonetheless, it is noticeable that the average execution time largely depends on the filter configuration (\autoref{fig:execFilters}). Consequently, in some occasions the CPU performed better than the GPU (e.g. CPU 'Gaussian blur' vs. GPU 'Gaussian blur, Sobel and Dilate'). The results show that with filter addition, the average execution time grows almost in proportion, the most likely reason for which is the implementation. In order to perform two filters, the component is called twice because the component interface exposes single filters, and not their combinations. To avoid this, pipelining should be implemented, however it would produce larger number of functions (all combinations) or reduce flexibility of a component. The implementation of the component for the FPGA is pipelined, so regardless of the filter size, one can notice the fairly even execution time of about $21.8ms$. For FPGA standards this is a lot, and with better implementation it can be fairly close to times from CPU and GPU. 

To fully stress the Image filtering component allocated on the CPU and the GPU, 8K image was also tested. As shown in \autoref{tab:filteringExec}, the Sobel filter executed on the GPU is on average 5.5 times faster (without any particular optimizations) than on the CPU. 

\subsubsection*{Object detection component}

The average execution time of the Object detection component is presented in \autoref{tab:detectionExec}. In all configurations, component allocated on the GPU outperformed the one allocated on the CPU. Strictly from the execution time perspective the GPU is more efficient in performing the algorithm, however, measurements have shown that using the same Haar classifier the CPU performed better in finding the pattern within the image (the OpenCV version used was 2.4.11).  

\begin{table}[h!]
	\centering
		\caption{Detection component execution times}
		\resizebox{0.75\columnwidth}{!}{%
		\begin{tabular}{ |l|l|l|l|l| }
		\hhline{*{1}{~}*{4}{|-}}
			\multicolumn{1}{c|}{} & Img. size & Avg.exec.t. [ms] \cellcolor{Gray} & SD ($\sigma$) & Var ($\sigma^{2}$) \\ \hline\hline				
			\multirow{2}{*}{\rotatebox[origin=c]{90}{GPU}} 	
			&	Small					& 6.7713 \cellcolor{Gray}	&	1.0287 &	1.0582 \\ 		
			& Big						&	16.2945	\cellcolor{Gray}&	2.2571 &	5.0946  \\ \hline \hline
			\multirow{2}{*}{\rotatebox[origin=c]{90}{CPU}} 
			&	Small					& 33.2952	\cellcolor{Gray}&	1.6500 		&	2.7225 \\ 
			& Big						&	192.9166 \cellcolor{Gray}&	3.2467 	&	10.5412 \\ \hline
			\multicolumn{5}{l}{\textsl{* big image: 1920x1080px, small image: 853x480px}}
		\end{tabular}
		}
	\label{tab:detectionExec}
\end{table}

\clearpage

\section{Related work}

There are numerous studies that compare the performance of CPUs, GPUs and FPGAs. Since the results of the performance are ambiguous in deciding which platform is the best, authors tend to focus on specific use cases, e.g. image processing, basic linear algebra subroutines, etc. 

A research by Asano et al. reports that for image processing (2D filters), a GPU is by far the best platform, followed by a CPU, but only for filters up to a certain size, beyond which an FPGA surpasses both a CPU and a GPU~\cite{Asano2009}. However, the experiment results by Pauwels et al., show that for real--time image processing an FPGA is without doubt the most suitable platform~\cite{Ros2012}. A comparison between a CPU and a GPU by Lee et al., showed that the performance gap between these platforms is not different in orders of magnitude as it is often considered~\cite{Lee2010}.

For software architects this means that the current software models must support multiple criteria. There are several papers that take into consideration software profiling from the perspective of energy consumption~\cite{Seo2007}, and a lot more from the perspective of execution time~\cite{Tsoi2011,Lee2006}, however additional research in this area is required for component based frameworks. 

\section{Conclusion and future work}

This paper presents the result of power consumption and execution time profiling for the software components that can be allocated on a heterogeneous computing platform. Solely based on the average power consumption, an FPGA seems to be the best choice to allocate the Image filtering component. Based on execution time, the GPU is the best choice for both Image filtering and Object detection components, in most cases.
In some occasions (\autoref{tab:filteringPower}), since it performs the tasks much faster, the GPU consumed on average less power than the CPU. But, since the upper limit of the power consumption of the CPU is much less than for the GPU, for larger tasks the GPU consumed more power than CPU. 
However, in cases with less data to handle, the CPU performs just as well. The CPU has proven to be the \textsl{all--round platform in the middle}, efficient from both power consumption and execution time perspective. 

Considering the software components in this paper, it is not resolved which platform is the absolute winner. Each one is well suited for a particular purpose, i.e. scenario. The software architect faced with the decision to allocate software components to a particular computing platform therefore needs: a) a scenario based decision approach, b) a decision support framework capable of handling multiple criteria. Consequently, the extra--functional properties of a component should reflect its internal implementation, because its performance is dependent on different scenarios. With a knowledge of usage scenario and the behavior of components in a particular scenario, software architects can make better design decisions.

Finally, the research presented in this paper will be continued in future focusing on: a) performing power consumption and execution time profiling for all the components, b) performing profiling for communication channels between the components, c) improving the measurement technique and d) creation of a decision support framework for guiding software architects faced with a challenge of allocating software components on heterogeneous computing environments.

%
\bibliographystyle{abbrv}
\bibliography{bibliography}  

\begin{thebibliography}{10}

\bibitem{Asano2009}
S.~Asano, T.~Maruyama, and Y.~Yamaguchi.
\newblock {Performance comparison of FPGA, GPU and CPU in image processing}.
\newblock {\em FPL 09: 19th International Conference on Field Programmable
  Logic and Applications}, pages 126--131, 2009.

\bibitem{Collange2009}
S.~Collange, D.~Defour, and A.~Tisserand.
\newblock {Power consumption of GPUs from a software perspective}.
\newblock {\em Lecture Notes in Computer Science}, 5544 LNCS(PART 1):914--923,
  2009.

\bibitem{Member2011}
I.~Crnkovic, S.~Sentilles, A.~Vulgarakis, and M.~R.~V. Chaudron.
\newblock {A Classification Framework for Software Component Models}.
\newblock {\em Software Engineering, IEEE Transactions on}, 37(5):593--615,
  2011.

\bibitem{Lee2006}
E.~a. Lee.
\newblock {Cyber-Physical Systems - Are Computing Foundations Adequate ?}
\newblock In {\em Position Paper for NSF Workshop On Cyber-Physical Systems:
  Research Motivation, Techniques and Roadmap}, volume~1, pages 1--9. Citeseer,
  2006.

\bibitem{Lee2010}
V.~W. Lee, P.~Hammarlund, R.~Singhal, P.~Dubey, C.~Kim, J.~Chhugani,
  M.~Deisher, D.~Kim, A.~D. Nguyen, N.~Satish, M.~Smelyanskiy, and
  S.~Chennupaty.
\newblock {Debunking the 100X GPU vs. CPU myth}.
\newblock {\em ACM SIGARCH Computer Architecture News}, 38(3):451, 2010.

\bibitem{Ros2012}
K.~Pauwels, M.~Tomasi, J.~{D\'{\i}az Alonso}, E.~Ros, and M.~M. {Van Hulle}.
\newblock {A Comparison of FPGA and GPU for real-time phase-based optical flow,
  stereo, and local image features}.
\newblock {\em IEEE Transactions on Computers}, 61(7):999--1012, 2012.

\bibitem{Poovendran2010}
R.~Poovendran.
\newblock {Cyber-physical systems: Close encounters between two parallel
  worlds}.
\newblock {\em Proceedings of the IEEE}, 98(8):1363--1366, 2010.

\bibitem{Seo2007}
C.~Seo, S.~Malek, and N.~Medvidovic.
\newblock {An energy consumption framework for distributed java-based systems}.
\newblock {\em Proceedings of the twenty-second \{IEEE/ACM\} international
  conference on Automated software engineering}, pages 421--424, 2007.

\bibitem{Tsoi2011}
K.~H. Tsoi and W.~Luk.
\newblock {Power profiling and optimization for heterogeneous multi-core
  systems}.
\newblock {\em ACM SIGARCH Computer Architecture News}, 39(4):8, 2011.

\bibitem{Svogor2013a}
I.~\v{S}vogor, I.~Crnkovi\'{c}, and N.~Vr\v{c}ek.
\newblock {An extended model for multi-criteria software component allocation
  on a heterogeneous embedded platform}.
\newblock {\em Journal of Computing and Information Technology},
  21(4):211--222, 2013.

\bibitem{Wolf2009}
W.~Wolf.
\newblock {Cyber-physical systems}.
\newblock {\em IEEE Computer}, 42(3):88--89, 2009.

\bibitem{Wolf2012}
W.~Wolf.
\newblock {\em {Computers as Components}}.
\newblock Elsevier, 1 edition, 2012.

\end{thebibliography}
%
%

\end{document}